\input{aipcheck.tex}

\documentclass[final,numbers]{aipproc}

\newcommand\doingARLO[2][]{%
  \ifx\mmref\undefined #1\else #2\fi
}

\layoutstyle{8x11double}
\begin{document}

\title {The effect of hyperfine mixing in electromagnetic and semileptonic decays of doubly heavy baryons}


\author{C. Albertus}{
  address={Departamento de F\'{\i}sica Fundamental e IUFFyM, Universidad de
  Salamanca, E-37008 Salamanca, Spain}}

\author{E. Hern\'andez}{
  address={Departamento de F\'{\i}sica Fundamental e IUFFyM, Universidad de
  Salamanca, E-37008 Salamanca, Spain},
}
\author{J. Nieves}{
  address={Instituto de F\'{\i}sica Corpuscular (IFIC), Centro Mixto
  CSIC-Universidad de Valencia, Institutos de Investigaci\'on de Paterna,
  Aptd. 22085, E-46071 Valencia, Spain},
}

\begin{abstract}We consider the role played by hyperfine mixing in electromagnetic and
semileptonic decays of doubly heavy baryons, which give rise to large
corrections to the decay widths in both cases. Our results
qualitatively agree with other calculations.
\end{abstract}

\date{\today}

\maketitle

\section{Introduction}

Due to heavy quark spin symmetry, in the infinite heavy quark mass
limit, the total spin of the heavy quark subsystem in a doubly heavy
baryon is well defined and can be set to $S_h=0,1$. This fact leads to
the most usual classification scheme of doubly heavy baryons. 
In Table~\ref{tab:clasif} we show the ground state $J^\pi=\frac12^+,\frac32^+$
doubly heavy baryons classified so that
$S_h$ is well defined, and to which we shall refer to as the
$S_h$-basis. Being ground states for the given quantum numbers a total
orbital angular momentum $L=0$ is naturally assumed.

\begin{table}
\caption{ Quantum numbers and quark content of ground-state doubly
heavy baryons. }\label{tab:clasif}
\vspace{-3mm}
\footnotesize
\begin{tabular}{cccc||cccc}
\hline
Baryon   & Quark content 
                      & $S_h$ 
                           & $J^\pi$  &Baryon         & Quark content 
                                                                & $S_h$ 
                                                                   & $J^\pi$\vspace{-.1cm}\\ 
            &(l=u,d)    &   &         &                &           &   &         \\ \hline
$\Xi_{cc}$   & \{c~c\}~l & 1 & 1/2$^+$ & $\Omega_{cc}$   & \{c~c\}~s & 1 & 1/2$^+$ \\ 
$\Xi_{cc}^*$ & \{c~c\}~l & 1 & 3/2$^+$ & $\Omega_{cc}^*$ & \{c~c\}~s & 1 & 3/2$^+$ \\ 
$\Xi_{bb}$   & \{b~b\}~l & 1 & 1/2$^+$ & $\Omega_{bb}$   & \{b~b\}~s & 1 & 1/2$^+$ \\ 
$\Xi_{bb}^*$ & \{b~b\}~l & 1 & 3/2$^+$ & $\Omega_{bb}^*$ & \{b~b\}~s & 1 & 3/2$^+$ \\ 
$\Xi_{bc}$   & \{b~c\}~l & 1 & 1/2$^+$ & $\Omega_{bc}$   & \{b~c\}~s & 1 & 1/2$^+$ \\ 
$\Xi_{bc}^*$ & \{b~c\}~l & 1 & 3/2$^+$ & $\Omega_{bc}^*$ & \{b~c\}~s & 1 & 3/2$^+$ \\ 
$\Xi_{bc}'$  & [b~c]~l   & 0 & 1/2$^+$ & $\Omega_{bc}'$  & [b~c]~s   & 0 & 1/2$^+$ \\ 
\hline
\end{tabular}%
\end{table}

Spin dependent terms in the quark-quark interaction are proportional
to the chromomagnetic moment of the heavy quark, therefore are ${\cal
O}(1/m_Q)$, and vanish in the limit of infinite heavy quark mass.  Due
to the finite value of the heavy quark masses, this hyperfine
interaction between the light and any of the heavy quarks can admix
both $S_h=0$ and $S_h=1$ spin components into the wave function. This
mixing is negligible for $cc$ and $bb$ baryons, as the antisymmetry of
the wave function would require higher orbital angular momenta or
radial excitations. On the other hand, in the $bc$ sector, one would
expect the role of the mixing to be noticeable and actual physical
states to be admixtures of the $B_{bc}$ and $B'_{bc}$ ($B$ = $\Xi$,
$\Omega$) states listed in Table~\ref{tab:clasif}. Physical states are
very close to the states where the spin of the light and the $c$ quark
couple to a well defined value $S_{qc}=0,1$. To these states we shall
call the $qc$-basis. In the $qc$-basis the effect of the hyperfine
mixing is minimized as corrections are proportional to the inverse of
the $b$ quark mass.

Masses are rather insensitive to the mixing, and most calculations
simply ignore the latter and work in the $S_h$ basis. However, mixing
has a dramatic effect in semileptonic and electromagnetic decays of
heavy baryons. Roberts and Pervin \citep{pervin1} pointed to this issue
for the very first time, and later confirmed that it actually has an
enormous impact in semileptonic transitions \citep{pervin2}.

We have investigated the role of hyperfine mixing in semileptonic and
electromagnetic decays in Refs.~\citep{albertus2} and \citep{albertus3},
qualitatively corroborating the findings of Roberts and Pervin for the
semileptonic case.

In this contribution we show results that clearly show the relevance
of hyperfine mixing to properly describe the semileptonic and
electromagnetic decay widths of doubly heavy baryons.  Our model uses
a variational procedure (see Ref.~\citep{albertus}) to determine the
masses and wave functions of the baryons in the $S_h$ basis. Mixing
coefficients and physical masses, obtained in Ref.~\citep{albertus2} by diagonalizing
the mass matrix, are shown in the next equation.
\begin{eqnarray*}
\label{eq:mix}
&&\hspace*{-.9cm}\Xi\,_{bc}^{(1)}=
\hspace{.3cm}0.902\,\Xi\,'_{bc}+0.431\,\Xi\,_{bc}\,,\ 
M_{\Xi\,_{bc}^{(1)}}=6967\,{\rm MeV},\nonumber\\
&&\hspace*{-.9cm}\Xi\,_{bc}^{(2)}= -0.431\,\Xi\,'_{bc}+0.902\,\Xi\,_{bc}\, ,\ 
M_{\Xi\,_{bc}^{(2)}}= 6919\,{\rm MeV},\nonumber\\
&&\hspace*{-.9cm}\Omega\,_{bc}^{(1)}=
\hspace{.25cm}0.899\,\Omega\,'_{bc}+0.437
\,\Omega\,_{bc}\,,\ 
M_{\Omega\,_{bc}^{(1)}}=7046\,{\rm MeV},\nonumber\\
&&\hspace*{-.9cm}\Omega\,_{bc}^{(2)}=
-0.437\,\Omega\,'_{bc}+0.899\,\Omega\,_{bc}\, ,\ 
M_{\Omega\,_{bc}^{(2)}}= 7005\,{\rm MeV}.
\end{eqnarray*}
\section{Semileptonic Decay}

Table~\ref{tab:widths} summarizes our results for semileptonic decay
widths in transitions involving only unmixed states. They are in fair
agreement with the relativistic calculation in Ref.~\citep{ebert2}, and
the agreement is also good with the results of Ref.~\citep{faessler}
but only for transitions involving $bc$ baryons.  Results in
Ref.~\citep{pervin2} are always smaller.  As seen in the table, the
values of the decay widths involving $B'_{bc}$ or $B_{bc}$ baryons
(with $B=\Omega,\Xi$) are very different. This fact anticipates that
the decay widths for transitions involving physical states could be
rather different. This is indeed the case as can be seen in
Table~\ref{tab:newwidths} where the results for mixed (physical)
states are shown. We thus confirm the findings of Ref.~\citep{pervin2}.
The actual values differ, a reflection of the
discrepancies already present in the unmixed case.

A very interesting feature is the fact that the decay width of the
$B^{(2)}_{bc} \to B_{cc}^*$ transition is much smaller than its
unmixed counterpart. This can be understood as follows: The
$B^{(2)}_{bc}$ physical state is very close to the $\hat{B}'_{bc}$
state of the $qc$ basis (see Ref.~\citep{albertus2}). In the latter,
$q$ and $c$ quarks couple to $S_{qc} = 0$, whereas in the $B^*_{cc}$
the light and any of the $c$ quarks couple to spin equal to one. In
this situation any spectator calculation, would find that the
amplitude of the transition $\hat{B}'_{bc} \to B_{cc}^*$ would cancel
due to orthogonality. The fact that $B^{(2)}_{bc}$ slightly deviates
from $\hat{B}'_{bc}$ produces a small but nonzero decay width.

\begin{table}
\caption{  Semileptonic decay widths  $({\rm in\ units \ of}\ 10^{-14}\ {\rm GeV})$ for unmixed states.
We use $|V_{cb}|=0.0413$.  $l=e,\mu$.} \label{tab:widths}
\vspace{-3mm}
\footnotesize
\begin{tabular}[tH]{lcccc||lcccc}
\hline
&\hspace*{-.5cm} This
work\hspace*{-.25cm}&\citep{ebert2}&\hspace*{-.25cm}\citep{faessler}&\hspace*{-.25cm}\citep{pervin2}
&&\hspace*{-.25cm}This
work\hspace*{-.25cm}&\citep{ebert2}&\hspace*{-.25cm}\citep{faessler}&\hspace*{-.25cm}\citep{pervin2}\\\hline
$\Gamma(\Xi_{bb}^*\to\Xi_{bc}'\,l\bar\nu_l)$ &  $1.08$  &$0.82$&
\hspace*{-.25cm}$0.36$&\hspace*{-.25cm}--
&$\Gamma(\Omega_{bb}^*\to\Omega_{bc}'\,l\bar\nu_l)$ & $1.14$
&$0.85$&\hspace*{-.25cm}$0.42$&\hspace*{-.25cm}-- \\ 
$\Gamma(\Xi_{bb}^*\to\Xi_{bc}\,l\bar\nu_l)$
&$0.36$&$0.28$&\hspace*{-.25cm}$0.14$&\hspace*{-.25cm}--
&$\Gamma(\Omega_{bb}^*\to\Omega_{bc}\,l\bar\nu_l)$
&$0.38$&$0.29$&\hspace*{-.25cm}$0.15$&\hspace*{-.25cm}--\\ 
$\Gamma(\Xi_{bb}\to\Xi_{bc}'\,l\bar\nu_l)$ &  $1.09$ &$0.82$&\hspace*{-.25cm}$0.43$&
\hspace*{-.25cm}$0.41$
&$\Gamma(\Omega_{bb}\to\Omega_{bc}'\,l\bar\nu_l)$ & $1.16$
&$0.83$&\hspace*{-.25cm}$0.48$&\hspace*{-.25cm}$0.51$ \\ 
$\Gamma(\Xi_{bb}\to\Xi_{bc}\,l\bar\nu_l)$ 
&$2.00$&$1.63$&\hspace*{-.25cm}$0.80$&\hspace*{-.25cm}$0.69$
&$\Gamma(\Omega_{bb}\to\Omega_{bc}\,l\bar\nu_l)$  &
$2.15$&$1.70$&\hspace*{-.25cm}$0.86$&\hspace*{-.25cm}$0.92$\\ 
$\Gamma(\Xi_{bc}'\to\Xi_{cc}\,l\bar\nu_l)$ & 
$1.36$&$0.88$&\hspace*{-.25cm}$1.10$&\hspace*{-.25cm}--
&$\Gamma(\Omega_{bc}'\to\Omega_{cc}\,l\bar\nu_l)$ &
$1.36$&$0.95$&\hspace*{-.25cm}$0.98$&\hspace*{-.25cm}-- \\ 
$\Gamma(\Xi_{bc}\to\Xi_{cc}\,l\bar\nu_l)$  &$ 2.57
$&$2.30$&\hspace*{-.25cm}$2.10$&\hspace*{-.25cm}$1.38$
&$\Gamma(\Omega_{bc}\to\Omega_{cc}\,l\bar\nu_l)$  &
$2.58$&$2.48$&\hspace*{-.25cm}$1.88$&\hspace*{-.25cm}$1.54$\\ 
$\Gamma(\Xi_{bc}'\to\Xi_{cc}^*\,l\bar\nu_l)$ & 
$2.35$&$1.70$&\hspace*{-.25cm}$2.01$&\hspace*{-.25cm}-- 
&$\Gamma(\Omega_{bc}'\to\Omega_{cc}^*\,l\bar\nu_l)$ & $2.35$
&$1.83$&\hspace*{-.25cm}$1.93$&\hspace*{-.25cm}--\\ 
$\Gamma(\Xi_{bc}\to\Xi_{cc}^*\,l\bar\nu_l)$  &$ 0.75 $ &$0.72$&\hspace*{-.25cm}$0.64$&\hspace*{-.25cm}$0.52$
&$\Gamma(\Omega_{bc}\to\Omega_{cc}^*\,l\bar\nu_l)$ 
&$0.76$&$0.74$&\hspace*{-.25cm}$0.62$&\hspace*{-.25cm}$0.56$\\
\hline
\end{tabular}%
\end{table}

\begin{table}
\caption{  Semileptonic decay widths  $({\rm in\ units\ of}\ 10^{-14}\ 
{\rm GeV})$ for mixed states.
}\label{tab:newwidths}
\vspace{-3mm}
\footnotesize
\begin{tabular}{lcc||lcc}
\hline
  &This work&\citep{pervin2}&&This work&\citep{pervin2}\\\hline
$\Gamma(\Xi_{bb}^*\to\Xi^{(1)}_{bc}\,l\bar\nu_l)$ &  $0.47$& --  
&$\Gamma(\Omega_{bb}^*\to\Omega^{(1)}_{bc}\,l\bar\nu_l)$ & $0.48$&--  \\ 
$\Gamma(\Xi_{bb}^*\to\Xi^{(2)}_{bc}\,l\bar\nu_l)$\ &$0.99$&--
&$\Gamma(\Omega_{bb}^*\to\Omega^{(2)}_{bc}\,l\bar\nu_l)$\ &$1.06$&--\\ 
$\Gamma(\Xi_{bb}\to\Xi^{(1)}_{bc}\,l\bar\nu_l)$ &  $2.21$& $0.95$ 
&$\Gamma(\Omega_{bb}\to\Omega^{(1)}_{bc}\,l\bar\nu_l)$ & $2.36$& $0.99$  \\ 
$\Gamma(\Xi_{bb}\to\Xi^{(2)}_{bc}\,l\bar\nu_l)$  &$0.85$& $0.33$
&$\Gamma(\Omega_{bb}\to\Omega^{(2)}_{bc}\,l\bar\nu_l)$  & $0.91$& $0.30$\\ 
$\Gamma(\Xi^{(1)}_{bc}\to\Xi_{cc}\,l\bar\nu_l)$ &  $0.38$& --
&$\Gamma(\Omega^{(1)}_{bc}\to\Omega_{cc}\,l\bar\nu_l)$ & $0.37$& -- \\ 
$\Gamma(\Xi^{(2)}_{bc}\to\Xi_{cc}\,l\bar\nu_l)$  &$ 3.50$&$ 1.92 $
&$\Gamma(\Omega^{(2)}_{bc}\to\Omega_{cc}\,l\bar\nu_l)$  & $3.52$& $1.99$\\ 
$\Gamma(\Xi^{(1)}_{bc}\to\Xi_{cc}^*\,l\bar\nu_l)$ &  $3.14$&--
&$\Gamma(\Omega^{(1)}_{bc}\to\Omega_{cc}^*\,l\bar\nu_l)$ & $3.14$&-- \\ 
$\Gamma(\Xi^{(2)}_{bc}\to\Xi_{cc}^*\,l\bar\nu_l)$  &$0.017$& $0.026$
&$\Gamma(\Omega^{(2)}_{bc}\to\Omega_{cc}^*\,l\bar\nu_l)$  &$0.014$&$0.013$\\
\hline
\end{tabular}%
\end{table}

\section{Electromagnetic Decay}

In Table~\ref{tab:gunmixeddiv} we show results for electromagnetic transitions
involving unmixed states.  We compare them with the results by 
Branz et al.~\citep{branz} obtained within a relativistic model.
Electromagnetic decay widths are proportional
to $(M_I-M_F)^3$, with $M_I,M_F$ the initial and final baryon masses,
showing a strong dependence on the actual baryon masses that are used. For that reason we show
results divided by that factor. In Table~\ref{tab:gdiv} we show the 
 results for physical states.  As for semileptonic decays we see that 
 the consideration of
hyperfine mixing has a tremendous effect on the decay widths.
In both cases the agreement with the results in    Ref.~\citep{branz} is poor.

\begin{table}\caption{Electromagnetic decay widths, divided by $(M_I-M_F)^3$, 
(in units of  $10^{-5}\ {\rm GeV^{-2}})$ for unmixed states.}
\label{tab:gunmixeddiv}
\begin{tabular}{lcc||lcc}
\hline
                                 & This work & Branz {\it et al.}    & 
                                 & This work &  Branz {\it et al.} \\ 

\hline
$\Xi_{bcu}^*\to\Xi'_{bcu}\gamma$   &  $73.6$    &     $57.0$         
& $\Omega_{bcs}^*\to\Omega'_{bcs}\gamma$    &  $72.8$   &   $58.3$   \\
$\Xi_{bcd}^*\to\Xi'_{bcd}\gamma$   &  $73.6$    &     $57.0$         &  
                                       &           &            \\\hline
$\Xi_{bcu}^*\to\Xi_{bcu}\gamma$    &  $333.9$   &       $471$       
 & $\Omega_{bcs}^*\to\Omega_{bcs}\gamma$     &  $87.7$   &   $160$    \\
$\Xi_{bcd}^*\to\Xi_{bcd}\gamma$    &  $160.6$   &       $231$    
    &                                         &           &            \\\hline
$\Xi_{bcu}'\to\Xi_{bcu}\gamma$     &  $36.7$    &       $57.8$  
     & $\Omega_{bcs}'\to\Omega_{bcs}\gamma$      &  $36.3$ & $57.4$    \\
$\Xi_{bcd}'\to\Xi_{bcd}\gamma$     &  $36.7$    &       $57.8$       & 
                                        &           &            \\\hline
\end{tabular}
\end{table}

\begin{table}
\caption{Electromagnetic decay widths, divided by $(M_I-M_F)^3$, (in units of 
 $10^{-5}\ {\rm GeV^{-2}})$ for mixed states.}\label{tab:gdiv}
\begin{tabular}{lcc||lcc}
\hline
                                        & This work & Branz {\it et al.}    &                                             & This work &  {\it Branz et al.} \\ 
\hline
$\Xi_{bcu}^*\to\Xi^{(1)}_{bcu}\gamma$       &  $248$  &   $293$              & $\Omega_{bcs}^*\to\Omega^{(1)}_{bcs}\gamma$    &  $12.7$   &  $0.8$     \\
$\Xi_{bcd}^*\to\Xi^{(1)}_{bcd}\gamma$       &  $4.9$  &   $0.39$             &                                             &          &             \\\hline
$\Xi_{bcu}^*\to\Xi^{(2)}_{bcu}\gamma$       &  $161.8$ &  $261$              & $\Omega_{bcs}^*\to\Omega^{(2)}_{bcs}\gamma$    &  $146$   &   $218$    \\
$\Xi_{bcd}^*\to\Xi^{(2)}_{bcd}\gamma$       &  $226$   &  $290$              &                                             &          &             \\\hline
$\Xi^{(1)}_{bcu}\to\Xi^{(2)}_{bcu}\gamma$    &  $112$   &  $126$              & $\Omega_{bcs}^{(1)}\to\Omega^{(2)}_{bcs}\gamma$ &  $124$   &   $215$     \\
$\Xi^{(1)}_{bcd}\to\Xi^{(2)}_{bcd}\gamma$    &  $189$   &  $280$              &                                             &          &             \\\hline
\end{tabular}
\end{table}

\section{Conclusions}
We qualitatively confirm the findings in Refs.~\citep{pervin1,pervin2}
as to the relevance of hyperfine mixing in $b\to c$ semileptonic
decays of doubly heavy baryons. Actual results differ by a factor of
two. We find mixing is also very important for electromagnetic decays
and, to our knowledge, our work in Ref.~\citep{albertus3} was the first
one where hyperfine mixing was considered for these transitions.
Electromagnetic decay widths are very sensitive to the actual baryon
masses and their precise determination demand accurate mass values.
Future experimental measurements of semileptonic and electromagnetic
decay widths will provide information on the mixing
parameters. However, this is not possible from the electromagnetic
transitions alone without relying in a theoretical model. The
situation is better in the case of semileptonic transitions where
first order heavy quark symmetry relations are accurate
enough \citep{albertus2}.

\section{Acknowledgments}

C. A. acknowledges a Juan de la Cierva contract from the Spanish
Ministry of Education. This research was supported by DGI and FEDER
funds, under contracts FIS2008-01143/FIS, FIS2006-03438,
FPA2007-65748, and the Spanish Consolider-Ingenio 2010 Programme CPAN
(CSD2007-00042), by Junta de Castilla y Le\'on under contracts SA016A07
and GR12, and by the EU HadronPhysics2 project, grant agreement
n. 227431.

\end{document}